\documentclass[12pt]{article}
\newcommand\be{\begin{equation}}
\newcommand\ee{\end{equation}}
\newcommand\bea{\begin{eqnarray}}
\newcommand\eea{\end{eqnarray}}
\newcommand\ket[1]{|#1\rangle}
\newcommand\bra[1]{\langle #1|}

\newcommand{\fatalpha}{{\bf \alpha \kern -0.44em \alpha}}
\newcommand{\fatsigma}{{\bf \sigma \kern -0.54em \sigma}}
\newcommand{\tpchi}{{\bf \chi \kern -0.35em \chi}}
\newcommand{\llambda}{{\bf \lambda \kern -0.45em \lambda}}

\bibliography{plain}
\title{\bf Quantum Correlation Dynamics for an Open Multi Qutrit System}\vspace{20mm}
\author{ R. Sufiani$^{a}$
  \thanks{E-mail:sofiani@tabrizu.ac.ir},
  A. Pedram
  \thanks{E-mail:alipedram01@gmail.com} and
 M. Karimi
 \thanks{E-mail:karimi-mohammad-70@yahoo.com}    ,
 \\ $^a$ {\small Department of Theoretical Physics and Astrophysics,
University of Tabriz, Tabriz 51664, Iran.} } \pagebreak

\vspace{20mm}
\usepackage{graphicx}
\usepackage{bookmark}
\usepackage{tensor}
\usepackage[fleqn]{amsmath}
\usepackage{float}
\begin{document}
\maketitle \vspace{15mm}
\newpage
\begin{abstract}
We study the correlation dynamics of a system composed of arbitrary numbers of qutrits interacting with a common environment. Initially, the system is assumed to be in a low dimensional subspace of the Hamiltonian called ``decoherence-free subspace". Environment induced quantum entanglement and discord is calculated between pair of qutrits of the system as measures of non-classical correlations. Finally the steady state distribution of entanglement and discord is determined with respect to the total number of qutrits of the system.

\end{abstract}
\vspace{70mm}
\newpage
\section{Introduction}
The defining characteristic of the quantum mechanics is existence of the correlations which cannot be explained using any local classical theory. Much work has been done to quantify these correlations, however it is not an easy task to give a general measure for "quantumness" in multidimensional systems. Entanglement and discord are two measures to capture the amount of non-classicality of a system. \\
Research in quantum information science has shown that entanglement and discord can be used as a resource in various communication protocols and search algorithms[\cite{1}, \cite{2}, \cite{3}, \cite{4}]. However, it is well known that entanglement is fragile under noisy processes while discord is more robust. In recent years it has been shown that environment can induce entanglement between subsystems of a system in various environmental settings[\cite{5}, \cite{6}, \cite{7}, \cite{8}]. It is also shown that discord can be amplified between two uncoupled qubits in a common environment[\cite{9}]. In [\cite{10}] the authors studied entanglement dynamics for qubits dissipating into a common environment.\\
Here, we study the time variation of entanglement and discord in a system composed of arbitrary number of qutrits dissipating into a common environment. Initially, the system is assumed to be in a subspace of hamiltonian to which application of the dissipation operator gives another member of this subspace. This subspace is called the decoherence free subspace. We find that pairwise entanglement and discord is created between an initially excited qutrit and a qutrit initially in the ground state due to interaction with a common environment.
\section{Correlation Measures}
In this work, we will use negativity as a measure for quantum entanglement and geometric quantum discord as a measure for non-classical correlations. The two concepts are explained in this section.
\subsection{Negativity}
Negativity is a measure of quantum entanglement which is easy to compute. The negativity of subsystem $A$ can be defined in terms of a density matrix $\rho$ as[\cite{11}]:
\begin{equation}\label{2-1}
N(\rho)=\frac{{\parallel\rho^{T_{A}}\parallel}_1-1}{2}.
\end{equation}
In this equation $\rho^{T_{A}}$ is the partial transpose of $\rho$ with respect to subsystem $A$ and ${\parallel\rho^{T_{A}}\parallel}_1$ is the trace norm or sum of the eigenvalues of the operator $\rho^{T_{A}}$.
\subsection{Geometric Quantum Discord}
Geometric quantum discord is defined as[\cite{12}]:
\begin{equation}\label{2-2}
D^{geo}(\rho)=\underset{\chi}{min}{\parallel\rho-\chi\parallel}^2,
\end{equation}
in which the minimum is taken over all the classical states $\chi$. Here, ${\parallel\rho-\chi\parallel}^2=tr(\rho-\chi)^2$ is the square norm in the Hilbert-Schmidt space.\\
In systems composed of qutrits, any bipartite state can be expanded as[\cite{13}]:
\begin{equation}\label{2-3}
\rho=\frac{1}{9}[I_3\otimes I_3+\sum_{i=1}^{8}x_i\lambda_i\otimes I_3+\sum_{j=1}^{8}y_jI_3\otimes\lambda_j+\sum_{i,j=1}^{8}t_{i,j}\lambda_i\otimes\lambda_j].
\end{equation}
In this equation $\lambda_i$s are the well known Gell-Mann matrices. $x$, $y$ and $t$ are defined as:
\begin{multline}\label{2-4}
x_i=\frac{3}{2}tr[\rho(\lambda_i\otimes I_3)]=\frac{3}{2}tr(\rho_A\lambda_i)\\
y_j=\frac{3}{2}tr[\rho(I_3\otimes\lambda_j)]=\frac{3}{2}tr(\rho_B\lambda_j)\\
T=t_{i,j}=\frac{9}{4}tr[\rho(\lambda_i\otimes\lambda_j)],\\
\end{multline}
in which $\rho_A$ and $\rho_B$ are the reduced density matrices.\\
Lower bound for the geometric quantum discord is given by:
\begin{equation}\label{2-5}
D(\rho)\geq tr(CC^t)-\sum_{i=3}^{3}\eta_i=\sum_{i=4}^{9}\eta_i,
\end{equation}
in which $\eta_i$s are the eigenvalues of $CC^t$ and C is given by:
\begin{multline}\label{2-6}
C=\left(
  \begin{array}{cc}
    \frac{1}{3} & \frac{2}{3\sqrt{3}}y^t \\
    \frac{2}{3\sqrt{3}}x & \frac{2}{9}T \\
  \end{array}
\right).\\
\end{multline}
\section{Dynamics}
The dynamics of a Markovian open quantum system can be described using the following master equation which is called the Lindblad equation[\cite{14}].
\begin{equation}\label{3-7}
\frac{d\rho(t)}{dt}=\sum_{k}[2L_k\rho(t)L^{\dagger}_k-\{\rho(t),L^{\dagger}_kL_k\}]\equiv D\rho(t).
\end{equation}
Here, the dissipation rate has been set equal to 1 for simplicity. $L_1$ and $L_2$ are the Lindblad operators. In the case of qutrits, these operators can be expressed as:
\begin{multline}\label{3-8}
L_1=\frac{1}{2}\sqrt{A_2}(\lambda_1+i\lambda_2)\\
L_2=\frac{1}{2}\sqrt{A_3}(\lambda_4+i\lambda_5),\\
\end{multline}
where, the matrix form of the Lindblad operators is given by:
\begin{multline}\label{3-9}
L_1=\sqrt{A_2}\left(
                \begin{array}{ccc}
                  0 & 1 & 0 \\
                  0 & 0 & 0 \\
                  0 & 0 & 0 \\
                \end{array}
              \right)\Rightarrow\L_1=\sqrt{A_2}\ket{0}\bra{1}\\
L_2=\sqrt{A_3}\left(
                \begin{array}{ccc}
                  0 & 0 & 1 \\
                  0 & 0 & 0 \\
                  0 & 0 & 0 \\
                \end{array}
              \right)\Rightarrow\L_2=\sqrt{A_3}\ket{0}\bra{2}.\\
\end{multline}
\section{The Model}
Let's assume that our system is composed of $n$ qutrits intracting with a common environment at zero temperature. Suppose that these qutrits have spontaneous emission. The transition probability from $\ket{1}$ to $\ket{0}$ is $A_2$ and from $\ket{2}$ to $\ket{0}$ is $A_3$. We assume that no transition occurs from $\ket{2}$ to $\ket{1}$.\\
\begin{figure}[H]
\centering
\includegraphics[width=60mm]{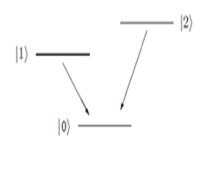}
\caption{A ``V" shaped 3-level atom \label{Fig.1}}
\end{figure}
Using the Lindblad operators we begin to solve the problem. We make use of decoherence free subspaces. First, we investigate the two-qutrit case. Assume that initially the system is in the state $\rho(0)=\ket{10}\bra{10}$. We apply the superoperator $D$ on this state and all of the subsequent states.
\begin{equation}\label{4-10}
D\ket{10}\bra{10}=2L\ket{10}\bra{10}L^{\dagger}-L^{\dagger}L\ket{10}\bra{10}-\ket{10}\bra{10}L^{\dagger}L,
\end{equation}
in which
\begin{multline}\label{4-11}
L=L_1\otimes I+I\otimes L_1+L_2\otimes I+I\otimes L_2\\
L^{\dagger}=L^{\dagger}_1\otimes I+I\otimes L^{\dagger}_1+L^{\dagger}_2\otimes I+I\otimes L^{\dagger}_2.\\
\end{multline}
Doing so, we get
\begin{multline}\label{4-12}
D\ket{10}\bra{10}=2A_2\ket{00}\bra{00}-2A_2\ket{10}\bra{10}-[\ket{10}(A_2\bra{01}+\sqrt{A_2A_3}(\bra{02}+\bra{20}))\\
+(A_2\ket{01}+\sqrt{A_2A_3}(\ket{02}+\ket{20}))\bra{10}]\\
D\ket{00}\bra{00}=0\\
D[(A_2\ket{01}+\sqrt{A_2A_3}(\ket{02}+\ket{20}))(A_2\bra{01}+\sqrt{A_2A_3}(\bra{02}+\bra{20}))]=\\
2A_2(A_2+2A_3)^2\ket{00}\bra{00}-2(A_2+2A_3)[(A_2\ket{01}+\sqrt{A_2A_3}(\ket{02}+\ket{20}))\\
(A_2\bra{01}+\sqrt{A_2A_3}(\bra{02}+\bra{20}))]-A_2(A_2+2A_3)[\ket{10}(A_2\bra{01}+\\
\sqrt{A_2A_3}(\bra{02}+\bra{20}))+(A_2\ket{01}+\sqrt{A_2A_3}(\ket{02}+\ket{20}))\bra{10}]\\
D[\ket{10}(A_2\bra{01}+\sqrt{A_2A_3}(\bra{02}+\bra{20}))+(A_2\ket{01}+\sqrt{A_2A_3}(\ket{02}+\ket{20}))\bra{10}]=\\
4A_2(A_2+2A_3)\ket{00}\bra{00}-2A_2(A_2+2A_3)\ket{10}\bra{10}\\
-2[(A_2\ket{01}+\sqrt{A_2A_3}(\ket{02}+\ket{20}))(A_2\bra{01}+\sqrt{A_2A_3}(\bra{02}+\bra{20}))]\\
-2(A_2+A_3)[\ket{10}(A_2\bra{01}+\sqrt{A_2A_3}(\bra{02}+\bra{20}))+(A_2\ket{01}+\sqrt{A_2A_3}(\ket{02}+\ket{20}))\bra{10}].\\
\end{multline}
Therefore, the decoherence free subspace is obtained as:
\begin{multline}\label{4-13}
H_{DFS}=span\{\ket{00}\bra{00},\ket{10}\bra{10},[(A_2\ket{01}+\sqrt{A_2A_3}(\ket{02}+\ket{20}))\\
(A_2\bra{01}+\sqrt{A_2A_3}(\bra{02}+\bra{20}))],[\ket{10}(A_2\bra{01}+\sqrt{A_2A_3}(\bra{02}+\bra{20}))+\\
(A_2\ket{01}+\sqrt{A_2A_3}(\ket{02}+\ket{20}))\bra{10}]\}.\\
\end{multline}
Now, we can expand the density matrix at time $t$ using these basis:
\begin{multline}\label{4-14}
\rho(t)=a_0(t)\ket{00}\bra{00}+a_1(t)\ket{10}\bra{10}+a_2(t)[(A_2\ket{01}+\sqrt{A_2A_3}(\ket{02}+\ket{20}))\\
(A_2\bra{01}+\sqrt{A_2A_3}(\bra{02}+\bra{20}))]+a_3(t)[\ket{10}(A_2\bra{01}+\sqrt{A_2A_3}(\bra{02}+\bra{20}))+\\
(A_2\ket{01}+\sqrt{A_2A_3}(\ket{02}+\ket{20}))\bra{10}]\}.\\
\end{multline}
Then, the following coupled equations are obtained:
\begin{multline}\label{4-15}
\dot{a}_0=2A_2a_1+2A_2(A_2+2A_3)^2a_2+4A_2(A_2+2A_3)a_3\\
\dot{a}_1=-2A_2a_1-2A_2(A_2+2A_3)a_3\\
\dot{a}_2=-2A_2(A_2+2A_3)a_2-2a_3\\
\dot{a}_3=-a_1-A_2(A_2+2A_3)a_2-2(A_2+A_3)a_3.\\
\end{multline}
Considering the initial conditions $a_0(0)=a_2(0)=a_3(0)=0,a_1(0)=1$, analytical solution of these equations will be as:
\begin{multline}\label{4-16}
a_0(t)=\frac{A_2}{2(A_2+A_3)}(1-e^{-4(A_2+A_3)t})\\
a_1(t)=\frac{A^2_2}{4(A_2+A_3)^2}(1+e^{-4(A_2+A_3)t})+\frac{A^2_2+A_2A_3}{2(A_2+A_3)^2}e^{-2(A_2+A_3)t}+(1-\frac{A_2}{A_2+A_3})\\
a_2(t)=\frac{1}{4(A_2+A_3)^2}(1+e^{-4(A_2+A_3)t})-\frac{1}{2(A_2+A_3)^2}e^{-2(A_2+A_3)t}\\
a_3(t)=\frac{A_2}{4(A_2+A_3)^2}(1+e^{-4(A_2+A_3)t})+\frac{A_3}{2(A_2+A_3)^2}e^{-2(A_2+A_3)t}-\frac{1}{2(A_2+A_3)}.\\
\end{multline}
Using these coefficients the density matrix can be determined at any arbitrary time.\\
For $n$ qutrits we assume that the initial state is $\rho(0)=\ket{k}\bra{k}$. By $\ket{k}$, we mean that the $k$th qutrit is in the excited state $\ket{1}$ and all of the other qutrits are in the ground state. Applying $D$ on the initial state and the subsequent states, we obtaain
\begin{multline}\label{4-17}
D\ket{k}\bra{k}=2A_2\ket{G}\bra{G}-2A_2\ket{k}\bra{k}-(\ket{E_{\not{k}}}\bra{k}+\ket{k}\bra{E_{\not{k}}})\\
D\ket{G}\bra{G}=0\\
D\ket{E_{\not{k}}}\bra{E_{\not{k}}}=2A_2((n-1)A_2+nA_3)^2\ket{G}\bra{G}-2((n-1)A_2+nA_3)\ket{E_{\not{k}}}\bra{E_{\not{k}}}\\
-A_2((n-1)A_2+nA_3)(\ket{E_{\not{k}}}\bra{k}+\ket{k}\bra{E_{\not{k}}})\\
D(\ket{E_{\not{k}}}\bra{k}+\ket{k}\bra{E_{\not{k}}})=4A_2((n-1)A_2+nA_3)\ket{G}\bra{G}-2A_2((n-1)A_2+nA_3)\ket{k}\bra{k}\\
-2\ket{E_{\not{k}}}\bra{E_{\not{k}}}-n(A_2+A_3)(\ket{E_{\not{k}}}\bra{k}+\ket{k}\bra{E_{\not{k}}})\\
\end{multline}
In these equations $\ket{G}$ corresponds to the state in which all qutrits are in ground state and
\begin{equation}\label{4-18}
\ket{E_{\not{k}}}:=A_2\sum_{i\not{=}k}^{n}\ket{i}+\sqrt{A_2A_3}\sum_{\mu=1}^{n}\ket{\mu}.\\
\end{equation}
Here, $\ket{i}$ and $\ket{\mu}$ are the states in which the $i$th and $\mu$th qutrit is in the excited states $\ket{1}$ and $\ket{2}$ respectively and all of the other qutrits are in ground state.\\
 Then, the corresponding decoherence free subspace is given by:
\begin{equation}\label{4-19}
H_{DFS}=span\{\ket{G}\bra{G},\ket{k}\bra{k},\ket{E_{\not{k}}}\bra{E_{\not{k}}},(\ket{E_{\not{k}}}\bra{k}+\ket{k}\bra{E_{\not{k}}})\}.
\end{equation}
The density matrix can be expanded in terms of the basis of the decoherence free subspace:
\begin{equation}\label{4-20}
\rho(t)=a_0(t)\ket{G}\bra{G}+a_1(t)\ket{k}\bra{k}+a_2(t)\ket{E_{\not{k}}}\bra{E_{\not{k}}}+a_3(t)(\ket{E_{\not{k}}}\bra{k}+\ket{k}\bra{E_{\not{k}}})
\end{equation}
Using (\ref{3-7}), we get this set of equations:
\begin{multline}\label{4-21}
\dot{a}_0=2A_2[a_1+((n-1)A_2+nA_3)^2a_2+2((n-1)A_2+nA_3)a_3\\
\dot{a}_1=-2A_2[a_1+((n-1)A_2+nA_3)a_3]\\
\dot{a}_2=-2((n-1)A_2+nA_3)a_2-2a_3\\
\dot{a}_3=-a_1-A_2((n-1)A_2+nA_3)a_2-n(A_2+A_3)a_3\\
\end{multline}
Solving the above equations, expressions for the coefficients are obtained as:
\begin{multline}\label{4-22}
a_0(t)=\frac{A^2_2}{n(A_2+A_3)}(1-e^{-2n(A_2+A_3)t})\\
a_1(t)=\frac{A_2}{n^2(A_2+A_3)^2}(1+e^{-2n(A_2+A_3)t})+\frac{2(n-1)A^2_2+2nA_2A_3}{n^2(A_2+A_3)^2}e^{-n(A_2+A_3)t}\\
+(1-\frac{2A_2}{n(A_2+A_3)})\\
a_2(t)=\frac{1}{n^2(A_2+A_3)^2}\{e^{-2n(A_2+A_3)t}-2e^{-n(A_2+A_3)t}+1\}\\
a_3(t)=\frac{A_2}{n^2(A_2+A_3)^2}(1+e^{-2n(A_2+A_3)t})+\frac{(n-2)A_2+nA_3}{n^2(A_2+A_3)^2}e^{-n(A_2+A_3)t}-\frac{1}{n(A_2+A_3)}.\\
\end{multline}
Then, the reduced density matrix of the 2-qutrit subsystem is given by:
\begin{multline}\label{4-23}
\rho_{k,l}(t)=[a_0(t)+\{(n-2)A_2(A_2+A_3)\}a_2(t)]\ket{00}\bra{00}+a_1(t)\ket{10}\bra{10}\\
+a_2(t)[(A_2\ket{01}+\sqrt{A_2A_3(\ket{02}+\ket{20})})+(A_2\bra{01}+\sqrt{A_2A_3(\bra{02}+\bra{20})})]\\
+a_3(t)[\ket{10}(A_2\bra{01}+\sqrt{A_2A_3}(\bra{02}+\bra{20}))+(A_2\ket{01}+\sqrt{A_2A_3}(\ket{02}+\ket{20}))\bra{10}]\\
\end{multline}
Now, we use negativity as the entanglement measure
\begin{equation}\label{4-24}
N(\rho_{k,l})=\frac{{\parallel\rho^{T_{k}}\parallel}_1-1}{2}.
\end{equation}
Taking partial trace with respect to the $k$th qutrit, we get:
\begin{multline}\label{4-25}
\rho^{T_{k}}(t)=[a_0(t)+\{(n-2)A_2(A_2+A_3)\}a_2(t)]\ket{00}\bra{00}+a_1(t)\ket{10}\bra{10}\\
+A_2a_3(t)[\ket{00}\bra{11}+\ket{11}\bra{00}]+A^2_2a_2(t)\ket{01}\bra{01}\\
+\sqrt{A_2A_3}a_3(t)[\ket{00}\bra{12}+\ket{12}\bra{00}+\ket{10}\bra{20}+\ket{20}\bra{10}]\\
+A_2\sqrt{A_2A_3}a_2(t)[\ket{00}\bra{21}+\ket{21}\bra{00}+\ket{01}\bra{02}+\ket{02}\bra{01}]\\
+A_2A_3a_2(t)[\ket{00}\bra{22}+\ket{22}\bra{00}+\ket{02}\bra{02}+\ket{20}\bra{20}].\\
\end{multline}
In Figure 2 the time variation of negativity is plotted for the two-qutrit case.
\begin{figure}[H]
\centering
\includegraphics[width=160mm]{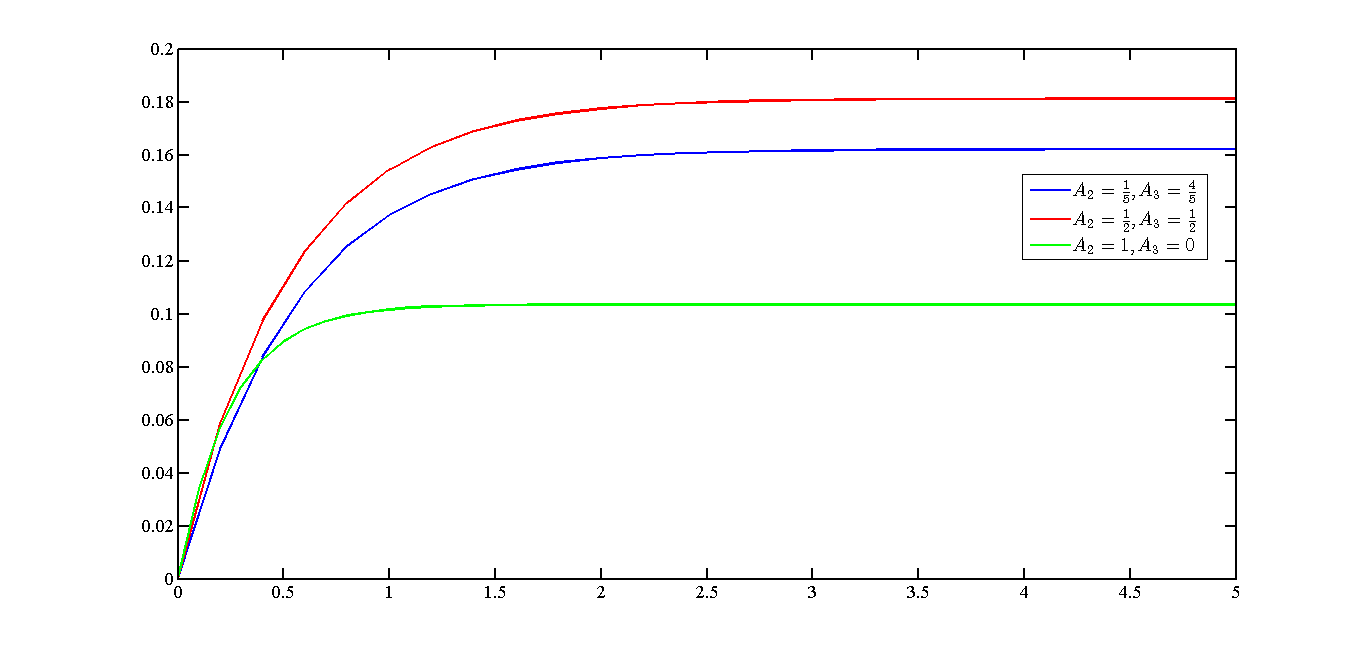}
\caption{Time variation of negativity for two-qutrit case using different $A_2$ and $A_3$ \label{Fig.2}}
\end{figure}
It is concluded that in a two-qutrit system with one qutrit in the ground state and the other in excited state, the interaction of system with environment induces entanglement between qutrits. The amount of entanglement reaches a steady value after a certain amount of time. Increasing the difference between $A_2$ and $A_3$ reduces the amount of entanglement and decreases the time at which entanglement reaches to the steady value.\\
Figure 3 shows the time variation of the negativity for multi-qutrit case.
\begin{figure}[H]
\centering
\includegraphics[width=160mm]{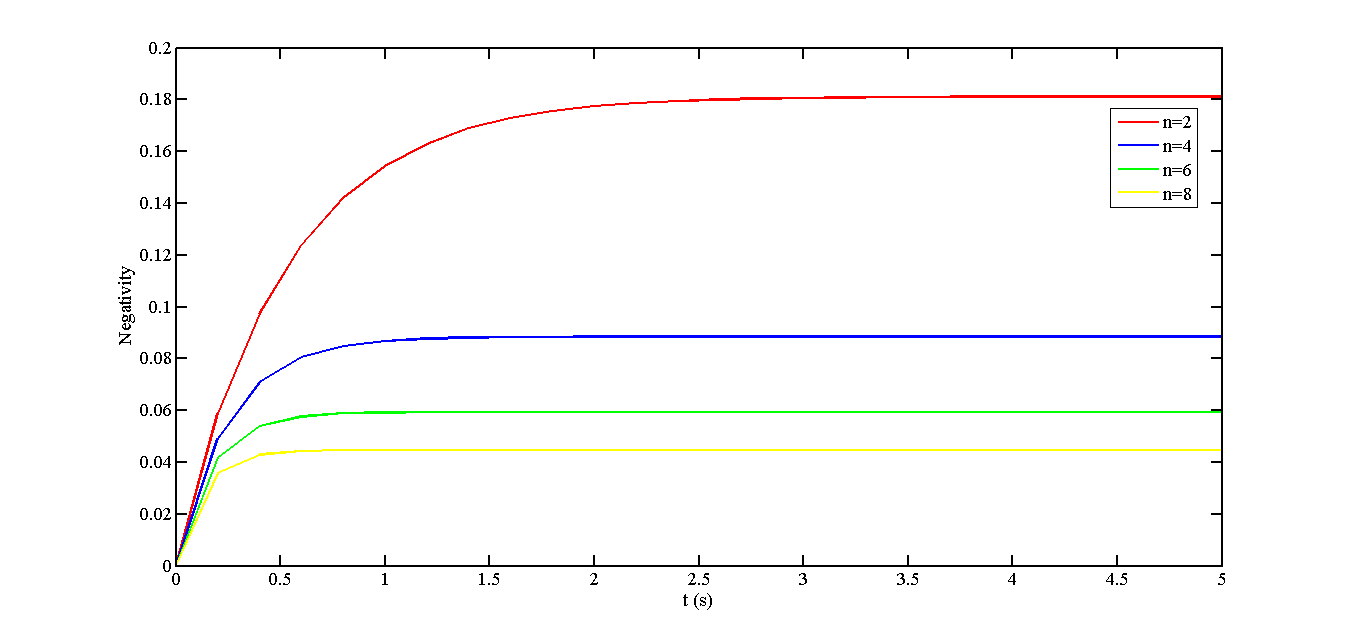}
\caption{Time variation of negativity for n-qutrit case \label{Fig.2}}
\end{figure}
We see that increasing the number of qutrits decreases the amount of entanglement and causes the system to reach its maximum negativity in a shorter time.
Now, we want to calculate the discord for a two-qutrit subsystem containing $k$th and $l$th qutrits. To do so, we will use the lower bound for quantum discord defined in equation (\ref{2-5}). We need to compute the matrix $C$ and therefore according to (\ref{2-6}) the matrices $x$, $y$ and $T$ should be calculated. To calculate the elements of these matrices we need to trace over the $k$th and $l$th qutrits in equation (\ref{4-23}).
\begin{multline}\label{4-26}
\rho_k(t)=tr_l(\rho_{k,l}(t))=[a_0(t)+a_1(t)+\{(n-2)A^2_2+(n-1)A_2A_3\}A_2(t)]\ket{0}\bra{0}\\
+A^2_2a_2(t)\ket{1}\bra{1}+A_2\sqrt{A_2A_3}a_2(t)(\ket{1}\bra{2}+\ket{2}\bra{1})+A_2A_3a_2(t)\ket{2}\bra{2}.\\
\end{multline}
Therefore, $x$s are obtained as:
\begin{multline}\label{4-27}
x_1=\frac{3}{2}tr(\rho_k\lambda_1)=0,\quad x_2=\frac{3}{2}tr(\rho_k\lambda_2)=0\\
x_3=\frac{3}{2}tr(\rho_k\lambda_3)=\frac{3}{2}[a_0(t)+a_1(t)+\{(n-3)A^2_2+(n-1)A_2A_3\}a_2(t)]\\
x_4=\frac{3}{2}tr(\rho_k\lambda_4)=0, \quad x_5=\frac{3}{2}tr(\rho_k\lambda_5)=0\\
x_6=\frac{3}{2}tr(\rho_k\lambda_6)=3A_2\sqrt{A_2A_3}a_2(t), \quad \quad x_7=\frac{3}{2}tr(\rho_k\lambda_7)=0\\
x_8=\frac{3}{2}tr(\rho_k\lambda_8)=\sqrt{\frac{3}{2}}[a_0(t)+a_1(t)+\{(n-1)A^2_2+(n-3)A_2A_3\}a_2(t)].\\
\end{multline}
Similarly:
\begin{multline}\label{4-28}
\rho_l(t)=tr_k(\rho_{k,l}(t))=[a_0(t)+\{(n-1)(A^2_2+A_2A_3)\}a_2(t)]\ket{0}\bra{0}\\
+a_1(t)\ket{1}\bra{1}+\sqrt{A_2A_3}a_3(t)[\ket{1}\bra{2}+\ket{2}\bra{1}]+A_2A_3a_2(t)\ket{2}\bra{2}.\\
\end{multline}
The expressions for $y$s are:
\begin{multline}\label{4-29}
y_1=\frac{3}{2}tr(\rho_1\lambda_1)=0,\quad y_2=\frac{3}{2}tr(\rho_1\lambda_2)=0\\
y_3=\frac{3}{2}tr(\rho_1\lambda_3)=\frac{3}{2}[a_0(t)-a_1(t)+\{(n-1)(A^2_2+A_2A_3)\}a_2(t)]\\
y_4=\frac{3}{2}tr(\rho_1\lambda_4)=0, \quad y_5=\frac{3}{2}tr(\rho_1\lambda_5)=0\\
y_6=\frac{3}{2}tr(\rho_1\lambda_6)=3\sqrt{A_2A_3}a_3(t), \quad y_7=\frac{3}{2}tr(\rho_1\lambda_7)=0\\
y_8=\frac{3}{2}tr(\rho_1\lambda_8)=\sqrt{\frac{3}{2}}[a_0(t)+a_1(t)+\{(n-2)A^2_2+(n-4)A_2A_3\}A_2(t)]\\
\end{multline}
The elements of matrix $T$ are also calculated according to equation (\ref{2-4}).\\
In Figure 4 the time variation of geometric quantum discord is plotted for the two-qutrit case.
\begin{figure}[H]
\centering
\includegraphics[width=160mm]{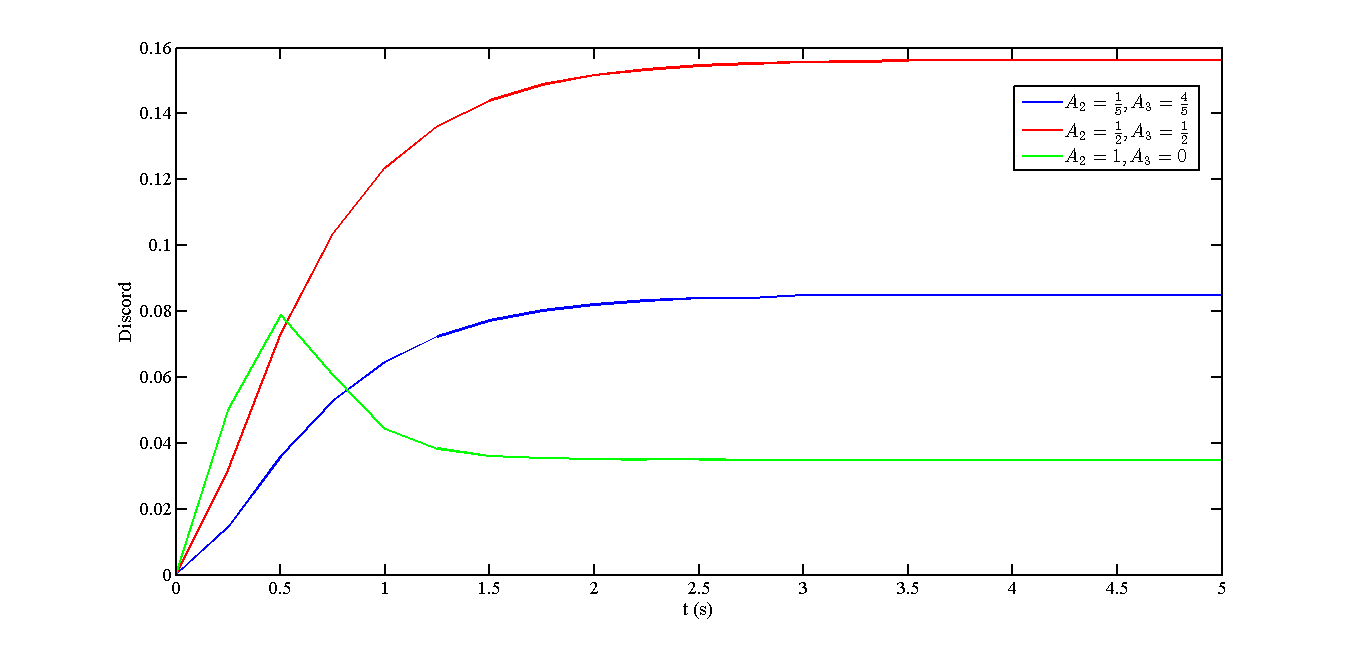}
\caption{Time variation of discord for 2-qutrit case \label{Fig.4}}
\end{figure}
The time variation of geometric quantum discord for multi-qutrit case is plotted in Figure 5.
\begin{figure}[H]
\centering
\includegraphics[width=160mm]{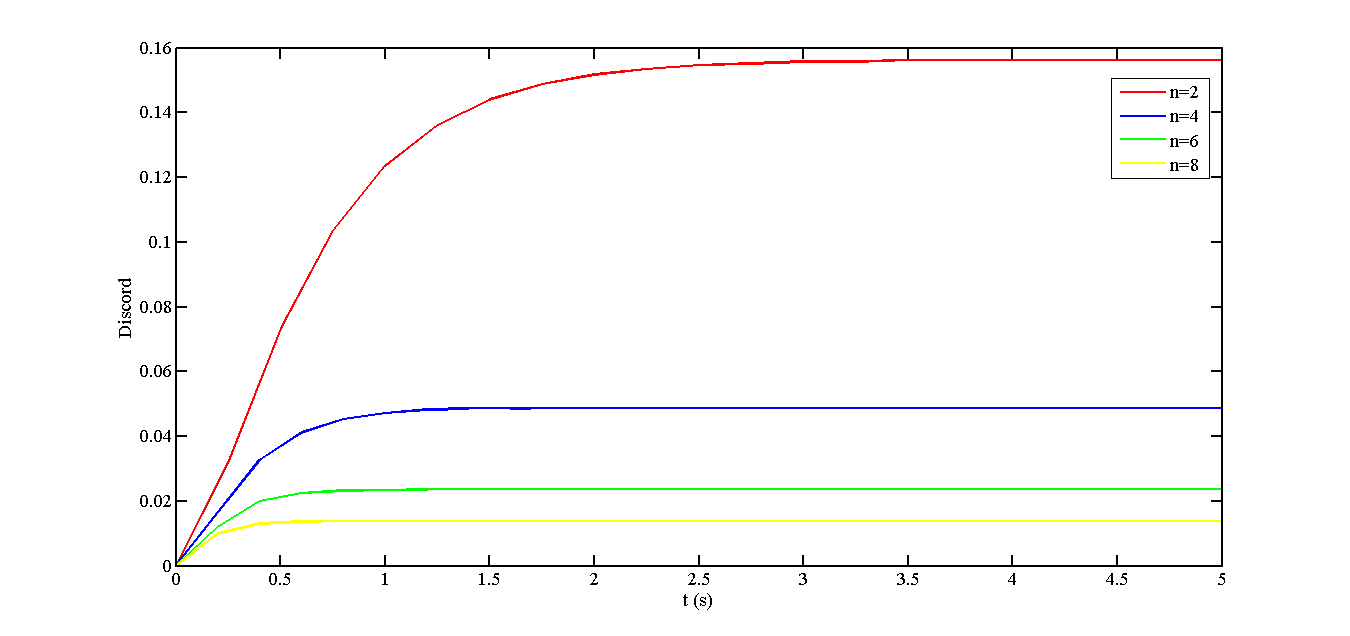}
\caption{Time variation of discord for n-qutrit case \label{Fig.5}}
\end{figure}
In figure (\ref{Fig.4}), we see that discord is generated between two initially uncorrelated qutrits. Increasing the difference between $A_2$ and $A_3$ results in decreased geometric quantum discord.\\
Figure (\ref{Fig.5}) shows that increasing the number of qutrits, decreases geometric quantum discord between the initially excited qutrit and the qutrits initially in the ground state. However, decreases the time at which discord reaches to the steady value.
\section{Conclusion}
In conclusion, we have studied the correlation dynamics of a system composed of arbitrary number of qutrits dissipating into a common environment. We found that quantum entanglement and discord is created and persists at steady state between two initially uncorrelated qutrits. Increasing the difference between the coefficients for spontaneous emission results in decreasing entanglement and discord. moreover, increasing the number of qutrits decreases the amount of entanglement and discord and causes the system to reach the steady state for these values in a shorter time.

\end{document}